\documentclass[aps,prd,preprint,nofootinbib,longbibliography]{revtex4-1}

\usepackage[utf8]{inputenc}
\usepackage{graphicx}
\usepackage{hyperref}
\usepackage{amsmath}
\usepackage{color}

\usepackage{etoolbox}
\apptocmd{\thebibliography}{\interlinepenalty=10000\relax}{}{}
\apptocmd{\sloppy}{\hbadness 10000\relax}{}{}

\AtBeginDocument{\count\footins=1000}

\providecommand{\color}[1]{}

\begin{document}

\title{Double Higgs boson production at \texorpdfstring{$e^+e^-$}{e+e-} colliders \texorpdfstring{\\}{}%
in the two-Higgs-doublet model}

\author{Tadashi Kon}
\email{kon@st.seikei.ac.jp}

\author{Takuto Nagura}
\email{dm186115@cc.seikei.ac.jp}

\author{Takahiro Ueda}
\email{tueda@st.seikei.ac.jp}

\author{Kei Yagyu\footnote{Address after February 2019:
Department of Physics, Osaka University, Toyonaka, Osaka 560-0043, Japan}}
\email{yagyu@het.phys.sci.osaka-u.ac.jp}

\affiliation{Department of Materials and Life Science,
Faculty of Science and Technology,
Seikei University,
3-3-1 Kichijoji-Kitamachi, Musashino-shi, Tokyo 180-8633, Japan}

\begin{abstract}
 
We study the double Higgs boson production processes $e^+e^- \to hh f\bar{f}$ ($f\neq t$) with $h$ being the 125 GeV Higgs boson in the two-Higgs-doublet model with a softly-broken $Z_2$ symmetry. 
The cross section can be significantly enhanced, typically a few hundreds percent, as compared to the standard model prediction due to resonant effects of heavy neutral Higgs bosons, which 
becomes important in the case without the alignment limit. 
We find a strong correlation between the enhancement factor of the cross section and the scaling factor of the $hf\bar{f}$ couplings
under constraints from perturbative unitarity, vacuum stability and current experimental data at the LHC as well as the electroweak precision data. 

 \end{abstract}
\maketitle

\section{Introduction}

Various signatures of the discovered Higgs boson at the LHC~\cite{Aad:2012tfa,Chatrchyan:2012xdj} show that its properties such as 
observed production cross sections and decay branching ratios are consistent with those predicted in the Standard Model (SM)~\cite{Khachatryan:2016vau}. 
Although the SM assumes the minimal form of the Higgs sector composed of one isospin scalar doublet, 
one may ask him- or herself a natural question: whether the discovered Higgs boson comes from just one doublet or not?
In fact, the discovered Higgs boson can be regarded as one of Higgs bosons arising from an extended structure of the Higgs sector, 
and there is no strong reason to restrict the Higgs sector to be minimal. 
On the other hand, extended Higgs sectors often appear as a low energy effective theory of physics beyond the SM based upon various physics motivations.
The important thing is that phenomenological features in the extended Higgs sectors strongly depend on a specific scenario of the underlying theory. 
Therefore, as a bottom-up approach, it is important to study the structure of the Higgs sector in order to narrow down new physics models. 

Among various possibilities of the extended Higgs sector, 
a two-Higgs-doublet model (THDM)~\cite{Branco:2011iw} is one of the simplest but important examples, as it appears in several new physics models. 
For example, models proposed to solve the gauge hierarchy problem predict THDMs as their low energy effective theories such as the 
minimal supersymmetric SM~\cite{Haber:1984rc} and composite Higgs models~\cite{Mrazek:2011iu,DeCurtis:2018iqd,DeCurtis:2018zvh}.  
In addition, extra CP violating phases can arise from the multi-doublet structure of the scalar potential, 
which are needed to realize the successful scenario for the electroweak baryogenesis~\cite{Turok:1990in,Nelson:1991ab}. 
Furthermore, the second Higgs doublet is often introduced in models beyond the SM to explain tiny neutrino masses~\cite{Zee:1980ai,Zee:1985rj,Ma:2006km,Campos:2017dgc,Camargo:2018uzw} and dark matter in the Universe~\cite{Barbieri:2006dq}. 
For these reasons, we consider the THDM  as a reference model, in which we 
impose a softly-broken discrete  $Z_2$ symmetry to avoid flavor changing neutral currents at tree level~\cite{Glashow:1976nt}. 

There are basically two ways to test the THDM at collider experiments, 
namely, the direct searches for additional Higgs bosons such as a charged Higgs boson and the indirect searches 
finding deviations in properties of the discovered Higgs boson ($h$) from the SM prediction. 
Concerning the former way, if we discover the additional Higgs bosons, it turns out to be a direct evidence for the THDM
or at least an extended Higgs sector, but 
no report has hitherto been provided for the discovery of such new particles at the LHC\@. 
Recent studies about the prospects of the direct searches at the high-luminosity LHC
(HL-LHC) include Refs.~\cite{Kling:2018xud,Adhikary:2018ise},
where possibilities to find new Higgs bosons via cascade decays of heavy scalar
resonances are discussed in detail.

Recently the latter way, seeking deviations as the indirect searches, is getting much attention,
since $h$ has already been discovered and its properties will be precisely measured in near future. 
For example, at the HL-LHC, the $h$ couplings to weak gauge bosons ($hVV$, $V=W,Z$) and fermions (e.g., $h\tau\tau$, $hb\bar{b}$) are expected to be measured with a few percent level~\cite{Dawson:2013bba}, while 
at the International Linear Collider (ILC), $hVV$ and $hf\bar{f}$ ($f = \tau,~c$ and $b$) couplings can be measured to be sub-percent and one percent level, respectively~\cite{Fujii:2017vwa}. 
In the THDM, these $h$ couplings can deviate from the SM prediction with various patterns depending on a particular scenario 
as it has been clarified in Ref.~\cite{Kanemura:2014bqa} at tree level and in Refs.~\cite{Arhrib:2003ph,Kanemura:2014dja,Kanemura:2017wtm,Kanemura:2015mxa,Kanemura:2018yai} at one-loop level. 
Therefore, by looking at the possible deviation in the $h$ couplings in the future collider experiments, 
one can distinguish the scenarios of the THDM,
which is still important even if an additional Higgs boson could be discovered
via a resonant peak in the direct searches.

In this paper, we focus on the cross section of double Higgs boson production processes at future $e^+e^-$ colliders 
as another important observable regarding the indirect search of the THDM\@. 
The double Higgs boson production has been discussed in the $pp$~\cite{Asakawa:2010xj,Grober:2017gut,Basler:2018dac}, $e^+e^-$~\cite{Asakawa:2010xj,Sonmez:2018smv} and $\gamma\gamma$~\cite{Asakawa:2010xj} collision in the THDM to extract the 
Higgs boson self-coupling constant $hhh$, particularly the case with the alignment limit, where all the SM-like Higgs boson couplings become the same as those in the SM prediction. 
In Ref.~\cite{Arhrib:2009hc}, the Higgs boson pair production has also been discussed at the LHC in the case with and without the alignment limit. 
Experimentally, it is a formidable task for the LHC to precisely determine
the Higgs trilinear coupling because of the tiny cross section of
the double Higgs boson production
(see \cite{ATLAS:2018otd,Sirunyan:2018two} for recent results).
This is certainly one of the motivations of construction of new powerful
colliders: the HL-LHC will be sensitive only to $\mathcal{O}(100\%)$ for
the trilinear coupling while the ILC running at 500 GeV can measure it
with an about 25\% uncertainty at the 68\% confidence level~\cite{DiVita:2017vrr}.

Our motivation to discuss the double Higgs production is to find the correlation between the deviation in the Higgs boson couplings and the modification of the cross section.  
These two variables are expected to be strongly correlated with each other, because 
the deviation in the $h$ couplings appears in the case without the alignment limit at tree level, in which case
the additional neutral Higgs bosons can mediate the double Higgs boson production process and can provide sizable enhancement of the cross section. 
For the LHC, large enhancement of the di-Higgs production cross section due to resonantly produced heavy Higgs bosons in the CP violating THDM
has been studied in Ref.~\cite{Grober:2017gut}.
In the present paper,
we clarify how large enhancement can be obtained at $e^+e^-$ colliders in the case without the alignment limit. 
In the numerical analysis, we focus on a special case of the THDM often referred to as Type-I, because 
scenarios without the alignment limit are highly constrained by
the Higgs boson signal strengths in the other types of Yukawa interactions such as Type-II\@. 
Under the current theoretical and experimental constraints on the parameter
space, we find a strong correlation between the enhancement of the cross section
and the scaling factor of the $h f \bar{f}$ couplings.
We note that Higgs boson pair production at $e^+e^-$ colliders requires
the collision energy $\sqrt{s}$ being \emph{larger} than 250~GeV while
the scaling factor can be precisely determined with $\sqrt{s} = 250~\text{GeV}$.
Hence, the Higgs boson couplings will have been  known when experiments reach to
the collision energy enough for Higgs pair boson production measurement.

This paper is organized as follows. 
In Sec.~\ref{sec:model}, we define the model and present the relevant Higgs boson interactions.
In Sec.~\ref{sec:hhX}, we discuss general properties of the Higgs boson pair production in $e^+e^-$ colliders.
Then, we explain a mechanism that enhances the cross section in the THDM without the alignment limit and give a rough estimate of the
enhancement factor.
Detailed numerical analysis on the aforementioned processes at
tree level is performed in Sec.~\ref{sec:numeric}.
Finally, Sec.~\ref{sec:conclusions} is devoted to our conclusions.


\section{Model}
\label{sec:model}

We consider the THDM whose Higgs sector is constructed by two isospin scalar doublet fields $\Phi_1$ and $\Phi_2$. 
For simplicity, we consider the CP-conserving case throughout the paper. 
The vacuum expectation values (VEVs) of the two doublets, i.e., $ \sqrt{2}\langle\Phi_{1,2}^0 \rangle = v_{1,2}$ are parameterized by 
$v = \sqrt{v_1^2 + v_2^2} = (\sqrt{2}G_F)^{-1/2}$ with $G_F$ being the Fermi constant and their ratio $\tan\beta = v_2/v_1$. 

It is convenient to introduce the so-called Higgs basis~\cite{Davidson:2005cw,Georgi:1978ri} as follows: 
\begin{align}
\begin{pmatrix}
\Phi_1 \\
\Phi_2
\end{pmatrix} = 
\begin{pmatrix}
\cos\beta & -\sin \beta \\
\sin \beta & \cos\beta 
\end{pmatrix}
\begin{pmatrix}
\Phi \\
\Phi'
\end{pmatrix}, 
\end{align}
where 
\begin{align}
\Phi = \begin{pmatrix}
G^+ \\
\frac{h_1' + v + i G^0}{\sqrt{2}}
\end{pmatrix},\quad 
\Phi' = \begin{pmatrix}
H^+ \\
\frac{h_2' + i A}{\sqrt{2}}
\end{pmatrix}. \label{HB}
\end{align}
In this basis, the Nambu-Goldstone bosons $G^\pm$ and $G^0$, which are absorbed into the longitudinal components of the $W$ and $Z$ bosons, are separated from 
the physical singly-charged Higgs boson $H^\pm$ and the CP-odd Higgs boson $A$, while two CP-even Higgs states $h_1'$ and $h_2'$ are generally not 
mass eigenstates at this stage. By introducing another mixing angle $\alpha$, the mass eigenstates are given by 
\begin{align}
\begin{pmatrix}
h_1' \\
h_2'
\end{pmatrix} = 
\begin{pmatrix}
c_{\beta-\alpha} & s_{\beta-\alpha} \\
-s_{\beta-\alpha} & c_{\beta-\alpha}
\end{pmatrix}
\begin{pmatrix}
H\\
h
\end{pmatrix}, 
\end{align}
where we have abbreviated $\sin X$ and $\cos X$ as $s_X$ and $c_X$, respectively. 
We identify $h$ as the discovered Higgs boson with a mass of about 125 GeV\@. 

The kinetic terms for the Higgs doublets are given as 
\begin{align}
{\cal L}_{\text{kin}} = \sum_{i=1,2}|D_\mu \Phi_i|^2  = |D_\mu \Phi|^2 + |D_\mu \Phi'|^2, \label{kin}
\end{align}
where $D_\mu$ is the covariant derivative for the Higgs doublets. 
We note that from the first term on the right-hand side, the gauge-gauge-Higgs type interactions are obtained as 
\begin{align}
{\cal L}_{\text{kin}} \supset \left(\frac{2m_W^2}{v} W_\mu^+ W^{-\mu}  + \frac{m_Z^2}{v} Z_\mu Z^\mu \right) (s_{\beta-\alpha}h +  c_{\beta-\alpha}H ). 
\end{align}
From the second term in the right-hand side of Eq.~(\ref{kin}), we obtain the Higgs-Higgs-gauge type interactions as 
\begin{align}
{\cal L}_{\text{kin}} & \supset -i\frac{g}{2}  H^+ \overleftrightarrow{\partial}^\mu   (c_{\beta-\alpha}h -  s_{\beta-\alpha}H - i A)W^-_\mu  + \text{h.c.}
- \frac{g}{2\cos\theta_W} A \overleftrightarrow{\partial}^\mu   (c_{\beta-\alpha}h -  s_{\beta-\alpha}H )Z_\mu\notag\\
&  - ie  H^+\overleftrightarrow{\partial}^\mu H^- A_\mu
- i\frac{g\cos2\theta_W}{2\cos\theta_W}   H^+\overleftrightarrow{\partial}^\mu H^- Z_\mu, 
\end{align}
with $X \overleftrightarrow{\partial}^\mu Y \equiv X (\partial^\mu Y) - (\partial^\mu X)  Y$ and $\theta_W$ being the weak mixing angle. 

In order to avoid Higgs boson mediating flavour changing neutral currents at tree level, we impose a discrete $Z_2$ symmetry~\cite{Glashow:1976nt} into the Higgs sector, where 
the two doublets are transformed as $\Phi_1 \to + \Phi_1$ and $\Phi_2 \to - \Phi_2$.  
Under the $Z_2$ symmetry, only one of the two Higgs doublets can couple to each up-type, down-type quarks and charged leptons, by which 
the interaction matrices in the flavour space between neutral Higgs bosons and fermions are diagonalized in the fermion mass eigenbasis. 
It has been known that there are four independent types of Yukawa interactions so-called Type-I, -II, -X and -Y~\cite{Aoki:2009ha} depending on 
the way to assign the $Z_2$ charge for fermions~\cite{Barger:1989fj,Grossman:1994jb}. 

The Yukawa Lagrangian for the third generation fermions is then given in the Higgs basis by 
\begin{align}
{\cal L}_Y &= -\bar{Q}_L^3 \frac{\sqrt{2}m_t}{v}\left( \tilde{\Phi} + \xi_t \tilde{\Phi}'  \right) t_R^{} 
-\bar{Q}_L^3  \frac{\sqrt{2}m_b}{v} \left(\Phi + \xi_b \Phi'  \right) b_R^{} \notag\\
& -\bar{L}_L^3 \frac{\sqrt{2}m_\tau}{v} \left(\Phi + \xi_\tau \Phi'  \right) \tau_R^{} + \text{h.c.},  \label{yukawa}
\end{align}
where $\tilde{\Phi}= i\tau_2\Phi^*$ and $\tilde{\Phi}' = i\tau_2\Phi^{\prime *}$. 
The factors $\xi_b$ and $\xi_\tau$ depend on the choice of the types of Yukawa interaction as 
\begin{align}
\begin{array}{ccccc}
& \text{Type-I} & \text{Type-II} & \text{Type-X} & \text{Type-Y}  \\
(\xi_b,\xi_\tau): & (\cot\beta, \cot\beta) & (-\tan\beta, -\tan\beta) & (\cot\beta, -\tan\beta) &  (-\tan\beta, \cot\beta), 
\end{array}
\end{align}
while $\xi_t = \cot\beta$ for all the types of Yukawa interaction. 
The interaction terms of the Higgs bosons and fermions are extracted as 
\begin{align}
{\cal L}_Y & \supset -\sum_{f = t,b,\tau}\frac{m_f}{v}\bar{f}\Big[ (s_{\beta-\alpha} + \xi_f c_{\beta-\alpha})h + (c_{\beta-\alpha} - \xi_f s_{\beta-\alpha})H  - 2iI_f \xi_f\, \gamma_5\, A\Big]f \notag\\
& - \frac{\sqrt{2}}{v} \bar{t}(m_bP_R - m_tP_L)H^+b - \frac{\sqrt{2}}{v} \bar{\nu}_\tau m_\tau P_R  H^+ \tau + \text{h.c.}, 
\end{align}
where $I_t\, (I_{b,\tau}) = 1/2\, (-1/2)$ and $P_L\, (P_R)$ is the projection operator for the left (right) hand chirality. 

The Higgs potential is generally written by eight independent real parameters when we include the soft-breaking term of the $Z_2$ symmetry: 
\begin{align}
V &= m_1^2 \Phi_1^\dagger\Phi_1 +  m_2^2 \Phi_2^\dagger\Phi_2 - m_3^2( \Phi_1^\dagger \Phi_2 + \text{h.c.})\notag\\
  & + \frac{\lambda_1}{2}(\Phi_1^\dagger \Phi_1)^2
    + \frac{\lambda_2}{2}(\Phi_2^\dagger \Phi_2)^2
    + \lambda_3 (\Phi_1^\dagger \Phi_1)(\Phi_2^\dagger \Phi_2)
    + \lambda_4 |\Phi_1^\dagger \Phi_2|^2
    + \frac{\lambda_5}{2}[(\Phi_1^\dagger \Phi_2)^2 + \text{h.c.}]. \label{pot}
\end{align}
As we already mentioned in the above, we assume the CP-invariance of the Higgs sector, so that $m_3^2$ and $\lambda_5$ parameters are taken to real. 
After imposing the tadpole conditions, i.e., 
the requirement of vanishing the linear terms of $h$ and $H$ in the Higgs potential, we can eliminate the $m_1^2$ and $m_2^2$ parameters. 
Then, these eight parameters, i.e., six parameters in the potential and two VEVs are expressed as follows: 
\begin{align}
m_{H^\pm},~~
m_{A}^{},~~
m_{H}^{},~~
m_{h},~~
M^2,~~
\tan\beta, ~~
v,~~
\alpha, \label{parameter}
\end{align}
where $M^2 \equiv m_3^2/(s_\beta c_\beta)$. 
Among these eight parameters, $v$ and $m_h$ are fixed to about 246 GeV and 125 GeV, respectively.  

From Eq.~(\ref{pot}), we can extract the scalar three-point couplings. 
In particular, the $hhh$ and $Hhh$ couplings will be important in the later analysis, which are expressed in terms of the parameters shown in (\ref{parameter}) as follows: 
\begin{align}
\lambda_{hhh}&=-\frac{m_h^2}{2v}s_{\beta-\alpha} +\frac{M^2-m_h^2}{v}s_{\beta-\alpha}c^2_{\beta-\alpha}+\frac{M^2-m_h^2}{2v}c_{\beta-\alpha}^3(\cot\beta-\tan\beta),  \label{lamhhh} \\
\lambda_{Hhh}&=-\frac{c_{\beta-\alpha}}{2v}\Big\{4M^2-2m_h^2-m_H^2 \notag\\
& \hspace{1cm}       +(2m_h^2+m_H^2-3M^2)[2c^2_{\beta-\alpha}+s_{\beta-\alpha}c_{\beta-\alpha}(\tan\beta-\cot\beta)]\Big\}, \label{lambhhh}
\end{align}
where the above quantities are defined by the coefficient in front of the $hhh$ and $Hhh$ vertices in the Lagrangian.

Before closing this section, it would be worth to mention that taking $s_{\beta-\alpha} \to 1$, all the $hVV$ ($V = W,Z$), $hf\bar{f}$ and $hhh$ couplings become the same values as in the SM at tree level. 
This limit has been known as the alignment limit, where the SM-like Higgs boson $h$ completely comes from the doublet $\Phi$ in the Higgs bases in Eq.~(\ref{HB}).

\section{Double Higgs boson production}
\label{sec:hhX}

\begin{figure}
\centering
\includegraphics[width=120mm]{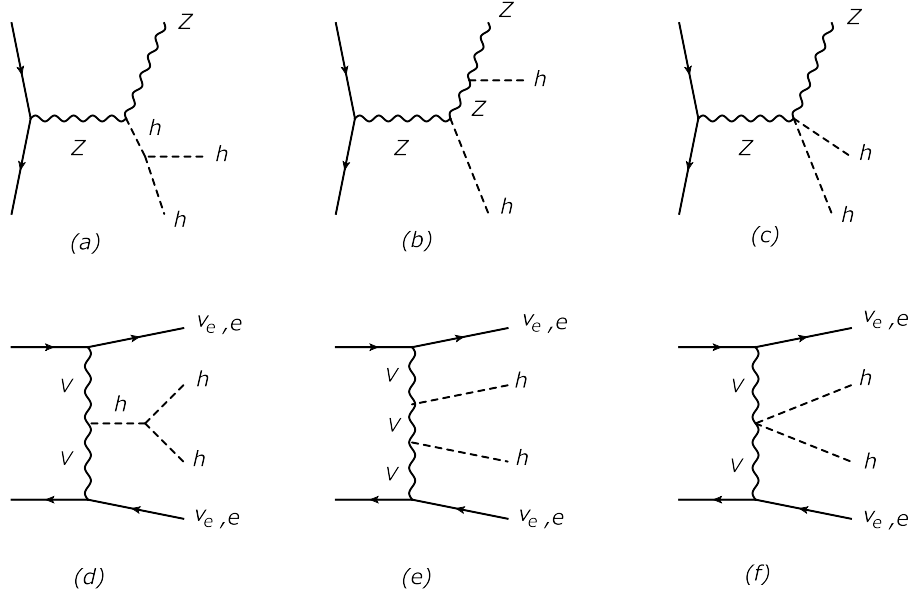}
\caption{Feynman diagrams for the double Higgs boson production process in the $e^+e^-$ collision in the SM with $V$ denoting either $W$ or $Z$. 
}
\label{diag1}
\end{figure}

\begin{figure}
\centering
\includegraphics[width=150mm]{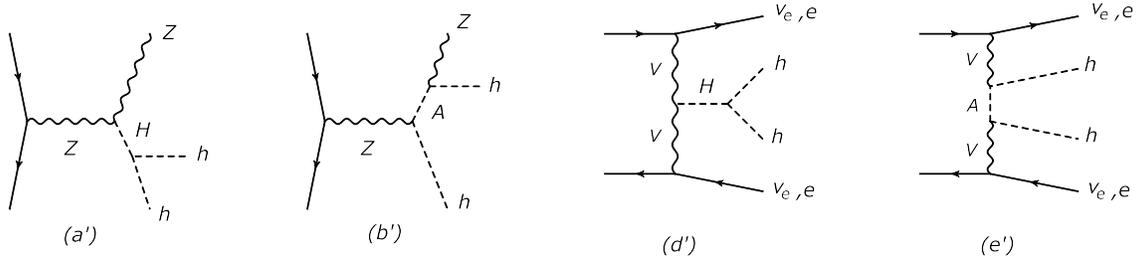}
\caption{Additional Feynman diagrams for the double Higgs boson production process in the $e^+e^-$ collision in the THDM with $V$ denoting either $W$ or $Z$. 
}
\label{diag2}
\end{figure}

In this section, we discuss the general property of double Higgs boson production processes at $e^+e^-$ colliders, and consider how much the production cross section can be 
different in the THDM with respect to that in the SM prediction. 

The double Higgs boson production $e^+e^- \to hh + X$ is possible when the collision energy of the electron and positron $\sqrt{s}$ is larger than about 250 GeV\@. 
Figs.~\ref{diag1} and \ref{diag2} show relevant Feynman diagrams, where those in Fig.~\ref{diag1} are common to the SM and THDM, while those in Fig.~\ref{diag2} only appear in the THDM\@. 
We here ignore diagrams induced via Yukawa couplings\footnote{For 
the case of $\sqrt{s}$ larger than $2(m_t + m_h) \sim 600$ GeV, the $t\bar{t}hh$ production is allowed and its contribution could be comparable as compared with that shown in Fig.~\ref{diag1}. 
In the numerical analysis of this paper given in the next section, we focus on the collision energy to be 500 GeV, so that this process is not needed to be considered. }. 
In the SM, there are two types of diagrams contributing to the double Higgs boson production, namely, the $s$-channel process $e^+e^- \to Zhh$ shown as diagrams (a)--(c) in Fig.~\ref{diag1} and the 
vector boson fusion processes $e^+e^- \to e^+e^-hh$ and $e^+e^- \to \nu_e \bar{\nu}_e hh$ shown as diagrams (d)--(f) in Fig.~\ref{diag1}. 
The cross section of these processes are calculated as 0.16~(0.12), $2.6 \times 10^{-3}$ ($7.2 \times 10^{-2}$) and $3.6 \times 10^{-4}$ ($9.5 \times 10^{-3}$)~fb 
for the $e^+e^- \to Zhh$, $\nu_e\bar{\nu}_ehh$ and $e^+e^-hh$ process at $\sqrt{s} = 500\ (1000)$ GeV, respectively. 
Therefore, the cross section of the double Higgs boson production is mainly determined by the $s$-channel process. 

In the THDM, additional diagrams contribute to the process as seen in Fig.~\ref{diag2}, where the extra neutral Higgs boson $H$ or $A$ appears in the diagrams having the same topology as 
those of (a), (b), (d) and (e) in Fig.~\ref{diag1}.  
In order to estimate the typical size of these contributions, let us focus on the diagram (a') in Fig.~\ref{diag2} as an example. 
When the mass of $H$ is taken to between $250~\text{GeV} \lesssim m_H^{} \lesssim \sqrt{s} -m_Z$, $H$ can be on-shell. 
In this case, the cross section of the diagram (a') is approximately calculated by the product of the two-body cross section of $e^+e^- \to ZH$ 
and the branching ratio of the $H \to hh$ decay assuming $\Gamma_H \ll m_H$ with $\Gamma_H$ being the total width of $H$.\footnote{In our scenario, the typical size of $\Gamma_H/m_H$ is 1\% level or smaller. } 
Therefore, the size of this cross section is typically obtained by multiplying the factor of $16\pi^2 \times c^2_{\beta-\alpha} \times \text{BR}(H \to hh)$ with respect to the cross section of 
the diagram (a) in the SM, where the factor $16\pi^2$ appears due to the typical ratio of the two-body and three-body phase-space factors, while $c^2_{\beta-\alpha}$ comes 
from the $HZZ$ coupling normalized to the $hZZ$ one in the SM\@. 
For example, when $s_{\beta-\alpha}$ is fixed to be 0.99 (0.995), the above factor becomes 3.14 (1.56) assuming BR$(H \to hh)= 1$. 
The similar enhancement can also be obtained from the diagram (b') as long as $A$ is produced with on-shell. 
We thus can expect that the total cross section of the double Higgs boson production can be several times larger than the SM prediction. 

The important thing here is that such enhancement of the cross section happens when departure of the alignment limit, i.e., $s_{\beta-\alpha} \neq 1$, is realized,  because the both 
$HZZ$ and $AZh$ couplings are proportional to $c_{\beta-\alpha}$. Therefore,  
the enhancement is strongly correlated with the deviation in the Higgs boson couplings from the SM prediction. 
On the other hand, the $hVV$ ($V=W,Z$) and $hf\bar{f}$ couplings are expected to be precisely measured at future $e^+e^-$ colliders. 
For example, at the ILC the $hZZ$, $hWW$, $hb\bar{b}$, $h\tau^+\tau^-$ and $hc\bar{c}$ couplings may be able to be measured with 
$0.38$, $1.8$, $1.8$, $1.9$ and $2.4$\% at 1$\sigma$ level assuming 250 GeV of the collision energy and 2 ab$^{-1}$ of the integrated luminosity~\cite{Fujii:2017vwa}.  
Therefore, if deviations in the Higgs boson couplings are detected in future, we expect the sizable enhancement of the double Higgs boson production as well. 

In the next section, we numerically evaluate how large enhancement can be obtained in the scenario without the alignment limit 
and show the correlation between the double Higgs boson production cross section and the 
deviation in the Higgs boson couplings.  

\section{Numerical results}
\label{sec:numeric}

In this section, we compute the cross section of the $e^+e^- \to f\bar{f} hh$ process, where $f$
is a fermion except for the top quark. These four-body final states are obtained through the $Zhh$
production with the decay of $Z$ into a fermion pair and the vector boson fusion process,
see Figs.~\ref{diag1} and~\ref{diag2}. 
We note that when the $Z$ boson from the $Zhh$ process decays into $e^+e^-$ ($\nu_e\bar{\nu}_e$), this
process interferes with the $e^+e^-hh$ $(\nu_e \bar{\nu}_ehh$) final states from the $Z$ ($W$) boson fusion process.
We take into account such interference effects in the numerical analysis. 
We also note that the dependence of the type of Yukawa interactions slightly appears in the cross section through
the decay width of $H$ and $A$. 
In contrast, the type dependence significantly appears in the region of the parameter space allowed from experimental bounds. 
In particular, except for the Type-I THDM, the scenario with $s_{\beta-\alpha} \neq 1$ which is considered in this section is
highly constrained by the Higgs boson signal strengths, see e.g.,~\cite{Blasi:2017zel}. 
We thus consider the Type-I THDM in what follows\footnote{In the Type-I THDM, 
constraints from flavor experiments such as those from $B \to X_s\gamma$ data on the charged Higgs boson mass are 
also quite milder than those in the other types especially in the Type-II\@. 
For example, ${\cal O}(100)$ GeV of the charged Higgs boson mass is allowed by the $B \to X_s\gamma$ data when $\tan\beta \gtrsim 2$~\cite{Misiak:2017bgg} in the Type-I THDM, 
while $m_{H^\pm} \lesssim 600$ GeV is excluded with 95\% confidence level in the Type-II THDM\@. }. 

As seen in (\ref{parameter}), there are six free parameters in the THDM, i.e., three masses of extra
Higgs bosons, $M^2$, $\tan\beta$ and $\alpha$. 
Instead of inputting $\alpha$, we take $s_{\beta-\alpha}$ and the sign of $c_{\beta-\alpha}$ as inputs. 
To avoid large contributions to the electroweak oblique $T$ parameter~\cite{Peskin:1990zt,Peskin:1991sw}, we take
the masses of $H^\pm$ and $A$ to be the same, i.e., $m_A^{} = m_{H^\pm}^{}$~\cite{Deshpande:1977rw,Sher:1988mj,Nie:1998yn,Kanemura:1999xf}. 
In this case, the quadratic dependence of the extra Higgs boson masses completely vanishes, and 
only the small logarithmic mass dependence remains, which is proportional to $c_{\beta-\alpha}^2$. 

Before going to show the numerical results for the cross section, let us summarize the constraints on the parameter space what we take into account in the analysis. 
We impose the perturbative unitarity bound~\cite{Kanemura:1993hm,Akeroyd:2000wc,Ginzburg:2003fe,Kanemura:2015ska} 
and the vacuum stability bound~\cite{Deshpande:1977rw,Sher:1988mj,Nie:1998yn,Kanemura:1999xf} as the constraints from theoretical consistency. 
These bounds restrict the size of scalar quartic couplings in the potential, which can be translated into the bound on the Higgs boson masses and the mixing angles. 
As the experimental constraints, we take into account the electroweak oblique $S$ and $T$ parameters~\cite{Tanabashi:2018oca},
direct searches for additional Higgs bosons at the LEP, Tevatron and LHC experiments as
well as the compatibility of the signal strengths for the discovered Higgs boson with a mass of 125 GeV\@. 
For the direct searches, we use the {\tt HiggsBounds-5.3.0beta}~\cite{Bechtle:2013wla}. 
For the signal strengths of the discovered Higgs boson, we use the combined data from ATLAS and CMS at the LHC
Run-I experiments~\cite{Khachatryan:2016vau}, and require that the prediction of the signal strengths for the $\gamma\gamma$, $ZZ^*$, $WW^*$ and $\tau\tau$
modes of the $h$ state does not exceed the given width of the error bar. We impose these
experimental bounds at the 95\% confidence level.

\begin{figure}
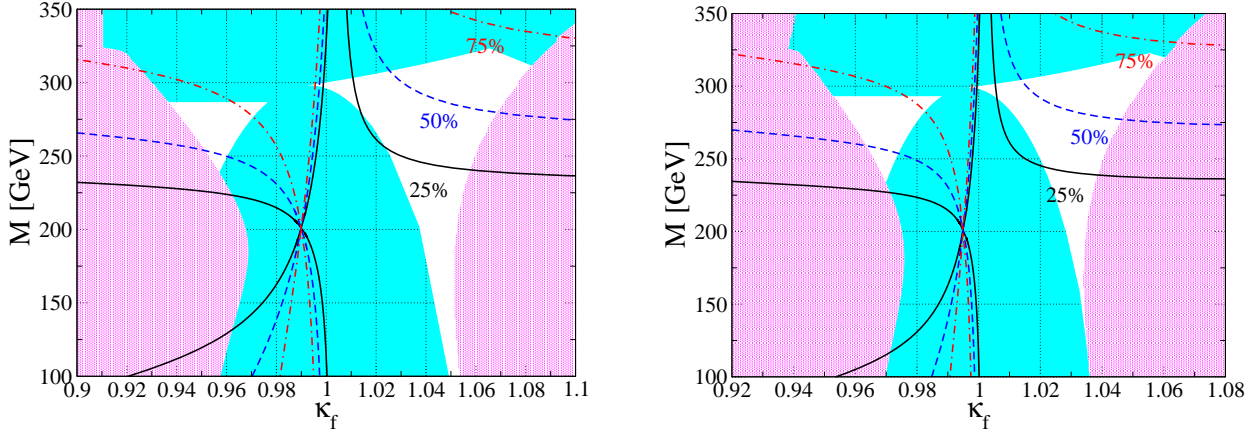

  \centering
  \includegraphics[width=0.47\textwidth]{bound_99.eps}
  \hfill
  \includegraphics[width=0.47\textwidth]{bound_995.eps}
  \caption{%
    Region excluded by the theoretical and experimental bounds in the case of
    $m_H^{} = m_A = m_{H^\pm}=300$ GeV and $s_{\beta-\alpha} = 0.99$ (left) and 0.995 (right).
    The blue (magenta) shaded region is excluded by the perturbative unitarity, vacuum stability bounds and/or the electroweak oblique $S$, $T$ parameters 
    (direct searches at the LEP, Tevatron and LHC experiments and/or compatibility of the signal strength of $h$). 
    The solid, dashed and dotted curves respectively show the contours for BR($H \to hh) = 0.25$, 0.5 and 0.75.
  }
  \label{const}
\end{figure}

In Fig.~\ref{const}, we show the region of the parameter space excluded by the theoretical and experimental bounds explained in the above. 
Here, we take $m_H^{} = m_A^{} (= m_{H^\pm}^{}) = 300$ GeV and $s_{\beta-\alpha} = 0.99$ (left panel) and 0.995 (right panel). 
The other parameters $M^2$ and $\tan\beta$ are scanned in these plots. 
Instead of showing the value of $\tan\beta$, we introduce the scaling factor $\kappa_f$ for the Yukawa coupling $g_{hff}^{\text{THDM}}$ in the THDM 
normalized to the SM value $g_{hff}^{\text{SM}}$, see Eq.~(\ref{yukawa}): 
\begin{align}
\kappa_f \equiv \frac{g_{hff}^{\text{THDM}}}{g_{hff}^{\text{SM}}} = s_{\beta-\alpha} + c_{\beta-\alpha} \cot\beta.  \label{kappaf}
\end{align}
We note that due to the choice of the Type-I THDM, the scaling factor $\kappa_f$ does not depend on the choice of a fermion $f$, see Sec.~\ref{sec:model}.  
From Eq.~(\ref{kappaf}), we see that $\kappa_f < s_{\beta-\alpha}$ ($\kappa_f > s_{\beta-\alpha}$) is obtained by taking the sign of $c_{\beta-\alpha}$ to be negative (positive), and 
$\kappa_f = s_{\beta-\alpha}$ is given in the limit of $\tan\beta \to \infty$. 

\begin{figure}
\centering
\includegraphics[width=0.7\textwidth]{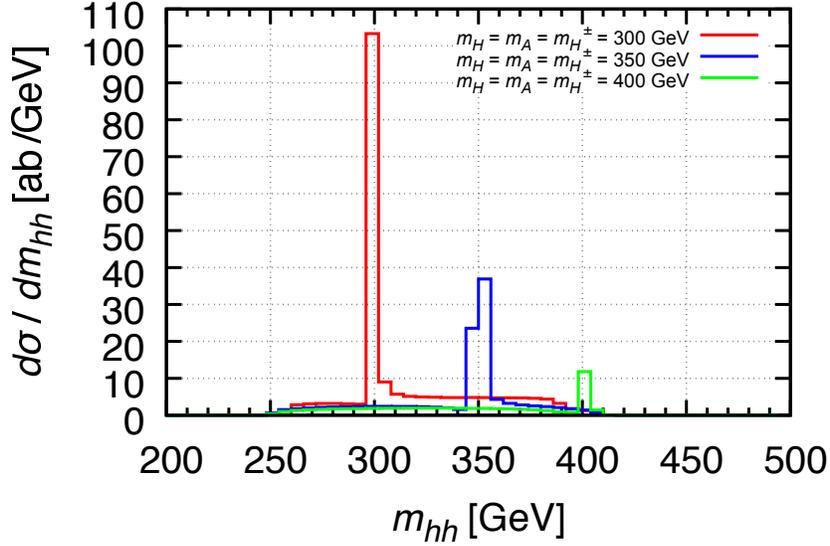}
\caption{Invariant mass distribution for the di-Higgs system $m_{hh}$. We take 
$s_{\beta- \alpha} = 0.99$, $c_{\beta-\alpha} > 0$ and $\tan \beta = 2$. 
For the red, blue and green histograms, we take $m_H = m_A = m_{H^\pm} = 300,~350$ and 400 GeV
with $M = m_H - 20$ GeV, respectively. All these choices of the parameters are allowed by the constraints explained in this section. }
\label{fig:inv}
\end{figure}

\begin{figure}
\centering
  \includegraphics[width=0.80\textwidth]{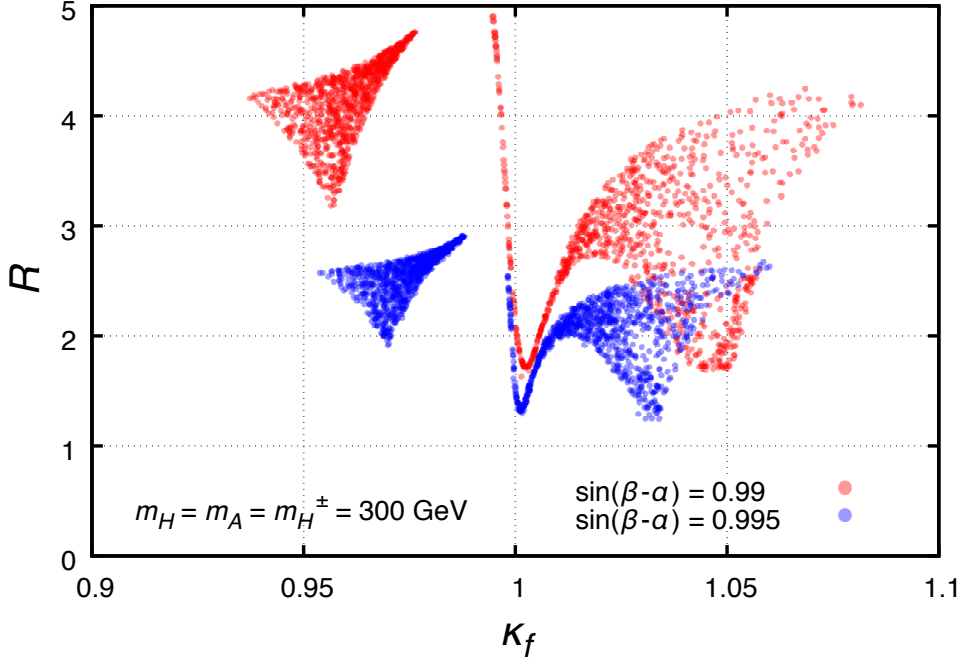}
  \caption{%
    Correlation between $\kappa_f$ and $R$ in the case of $m_H = m_A = m_{H^\pm} = 300$ GeV\@.
    We take $s_{\beta-\alpha} = 0.99$ (red points) and 0.995 (blue points).
    All the points satisfy the theoretical constraints and experimental bounds explained in this section.
  }
  \label{fig:mh300}
\end{figure}

\begin{figure}
  \centering

  \includegraphics[width=0.47\textwidth]{bound_99_500.eps}
  \hfill
  \includegraphics[width=0.47\textwidth]{bound_995_500.eps}
  \caption{%
    Same as in Fig.~\ref{const} but for the case of larger masses
    $m_A = m_{H^\pm} = 500\ \text{GeV}$ while the extra CP-even neutral Higgs
    mass is retained, $m_H = 300\ \text{GeV}$.
  }
  \label{const500}

  \vspace*{\floatsep}

  \includegraphics[width=0.80\textwidth]{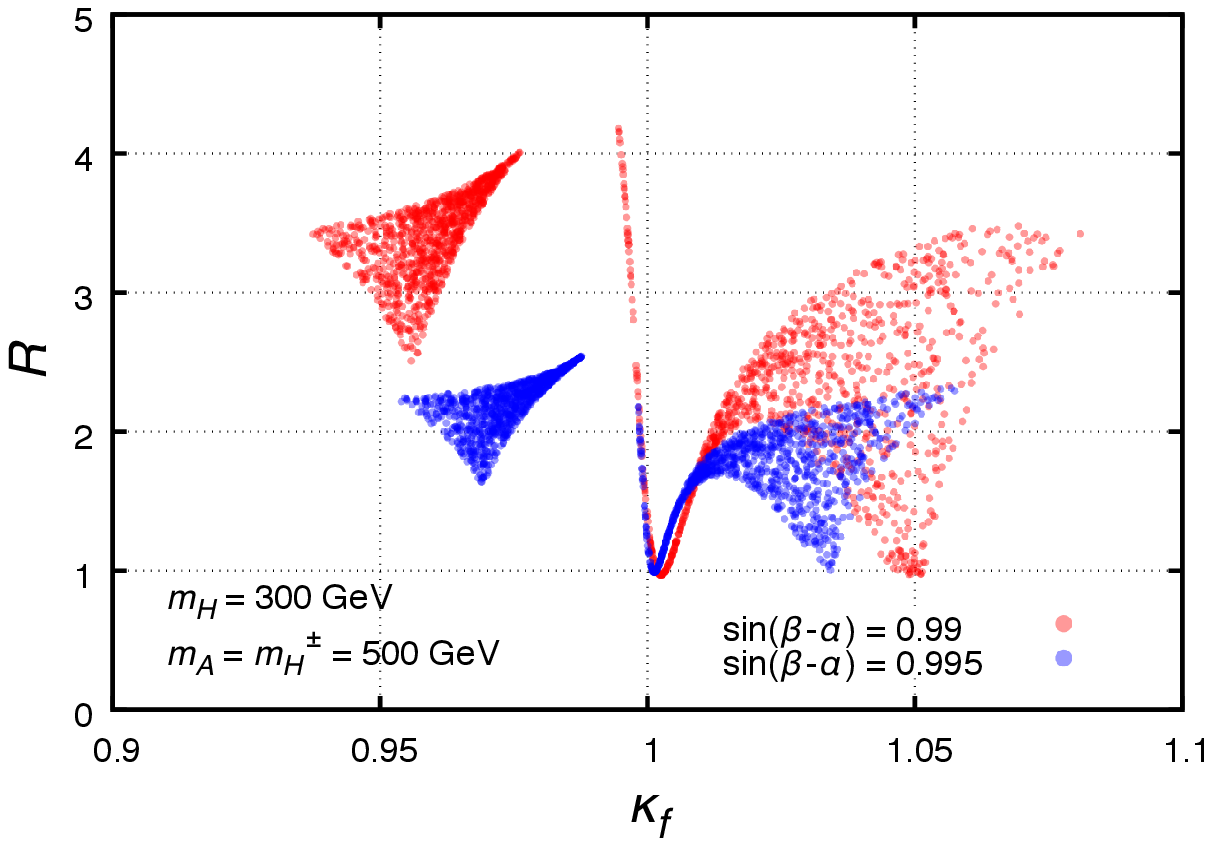}
  \caption{%
    Same as in Fig.~\ref{fig:mh300} but for the case of larger masses
    $m_A = m_{H^\pm} = 500\ \text{GeV}$ while the extra CP-even neutral Higgs
    mass is retained, $m_H = 300\ \text{GeV}$.
  }
  \label{fig:mA500}
\end{figure}

We see that the constraint from experiments (shown by the magenta shaded region) becomes important in the region with smaller $\tan\beta$ values, i.e., larger values of $|1-\kappa_f|$, where 
the direct search at the LHC, particularly for $H \to ZZ$ dominantly contributes to the exclusion.  
This can be understood by the fact that the production cross section $gg \to H$ is proportional to $\cot^2\beta$ in the limit of $s_{\beta-\alpha} \to 1$, 
so that the larger $\tan\beta$ case can avoid the constraint due to the smaller cross section. 
Another thing we can find in this figure is that the theory bound (shown by the blue shaded region\footnote{In fact, the constraints from the $S$ and $T$ parameter are also taken into account in the
blue shaded region, but the shape of the exclusion is almost determined by the perturbative unitarity and vacuum stability bounds. }) 
becomes important in the case with larger $\tan\beta$ and/or larger difference between $m_H^2$ and $M^2$. 
In particular, for a larger $\tan\beta$ case, quite small area with $M^2 \sim m_H^2$  is allowed, because some of the scalar quartic couplings become very sensitive to the value of $m_H^2-M^2$. 
The typical behavior of the constraints does not change so much between the cases with $s_{\beta-\alpha}=0.99$ (left panel) and 0.995 (right panel), 
but smaller values of $|1 - \kappa_f|$ are excluded by experimental constraints.  
This is simply because the value of $|1-\kappa_f|$ given by the same value of $\tan\beta$ becomes smaller in the case of $s_{\beta-\alpha} = 0.995$ as compared to the case of $s_{\beta-\alpha} = 0.99$, 
as seen in Eq.~(\ref{kappaf}). 
For reference, we also show the contours of the branching ratio of the $H \to hh$ mode which is important to understand the behavior of the enhancement of the cross section of the double Higgs boson production. 

Now we present numerical results on the cross section.
For the numerical computation of the cross section, we used the public version of {\tt GRACE}~\cite{Yuasa:1999rg,Fujimoto:2002sj,GRACE-2.2.1}
with some modifications, and all the calculations were performed at tree level with $\sqrt{s} = 500\ \text{GeV}$. 
Let us first illustrate the invariant mass distribution of the di-Higgs system $m_{hh}$ in Fig.~\ref{fig:inv}. 
Here, we take $m_H^{} = 300$, 350 and 400 GeV as examples, and the other parameters are specified as described in the caption.  
It is clearly seen that the sharp peaks appear at around $m_H^{}$, because of the on-shell $H$ mediation with the $H \to hh$ decay shown in the diagram ($a'$) of Fig.~\ref{diag2}. 
The height of the peak is getting lower as the mass of $H$ increases. 
Due to this resonant effect, the cross section of the double Higgs boson production is sizably enhanced as we will see in what follows. 

Next, we consider the ratio of the di-Higgs production cross section in the Type-I THDM to that in the SM in order to see how large enhancement can be obtained:
\begin{equation}
  R \equiv \frac{\sum_f\sigma^\text{THDM}(e^+ e^- \to f \bar{f} hh)}%
                {\sum_f\sigma^\text{SM}(e^+ e^- \to f \bar{f} hh)}, 
  \label{eq:R}
\end{equation}
where the summation for $f$ is done over all the fermions except for $f = t$. In the following calculation, we scan $\tan\beta$ and $M^2$ with the ranges of 
$1\leq\tan\beta \leq 30$ and $0\leq M^2 \leq (500~\text{GeV})^2$. 
The scatter plot in Fig.~\ref{fig:mh300} shows the ratio $R$ as functions of
the scaling factor $\kappa_f$ with using sampling points passing all the constraints, namely, points in the white region of Fig.~\ref{const}.
Red and blue points represent for $s_{\beta-\alpha} = 0.99$ and 0.995, respectively. 
As it is expected in Sec.~\ref{sec:hhX}, a clear correlation between $R$ and $\kappa_f$ is seen from this plot, and a considerable enhancement of the cross section is observed 
due to the on-shell mediation of the extra neutral Higgs bosons $H$ and $A$. See also the contours of the branching ratio of $H\to hh$ shown in Fig.~\ref{const} to figure out the 
behavior of the correlation.  
It is also seen that a larger value of $R$ is obtained in the case of $s_{\beta-\alpha} = 0.99$ in comparison with the case of $s_{\beta-\alpha} = 0.995$, because 
the $HZZ$ and $AZh$ couplings are proportional to $c_{\beta-\alpha}$. 
The strong enhancement seen in the large $\tan\beta$ region (i.e., $\kappa_f \sim s_{\beta-\alpha} \sim 1$) can be traced back to a large value of the $\lambda_{Hhh}$ coupling, 
see Eq.~(\ref{lambhhh}). 
This gives a considerable deviation $R>1$ even at $\kappa_f = 1$. 
We find that the value of $R$ can be maximally around 5 (3) for $s_{\beta-\alpha} = 0.99$ (0.995). 

We also perform the similar calculation shown in Figs.~\ref{const500} and \ref{fig:mA500}, but for the case with $m_H^{} = 300$ GeV and $m_A^{} = m_{H^\pm}^{} = 500$ GeV\@. 
The region allowed by the constraints is almost the same as the previous plots in Fig.~\ref{const}. 
On the other hand, by looking at Fig.~\ref{fig:mA500}, we find that the shape of the scattering points are shifted to below due to the $A$ mediation being off-shell, 
and the maximally allowed value of $R$ becomes around 4 and 2.5 for $s_{\beta-\alpha} = 0.99$ and 0.995, respectively.

\begin{figure}
\centering
\includegraphics[width=0.45\textwidth]{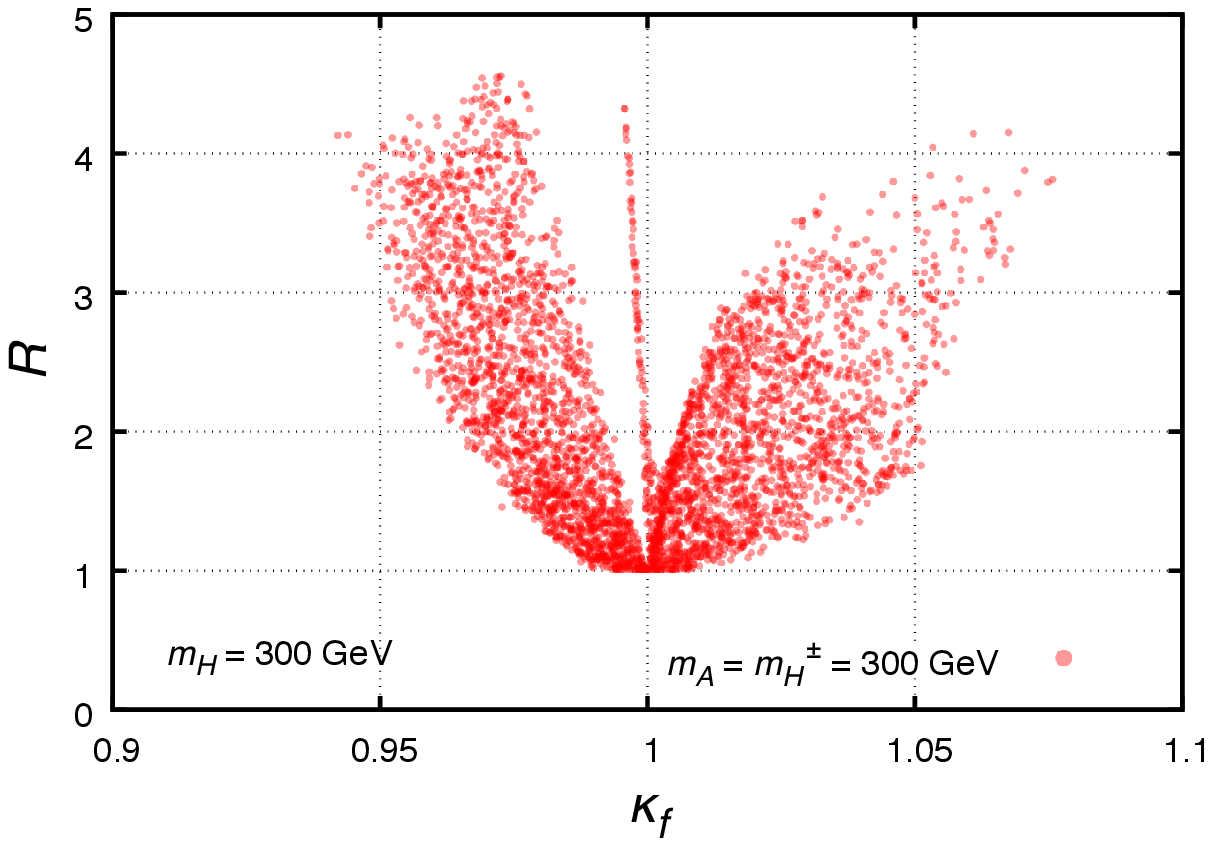}
\includegraphics[width=0.45\textwidth]{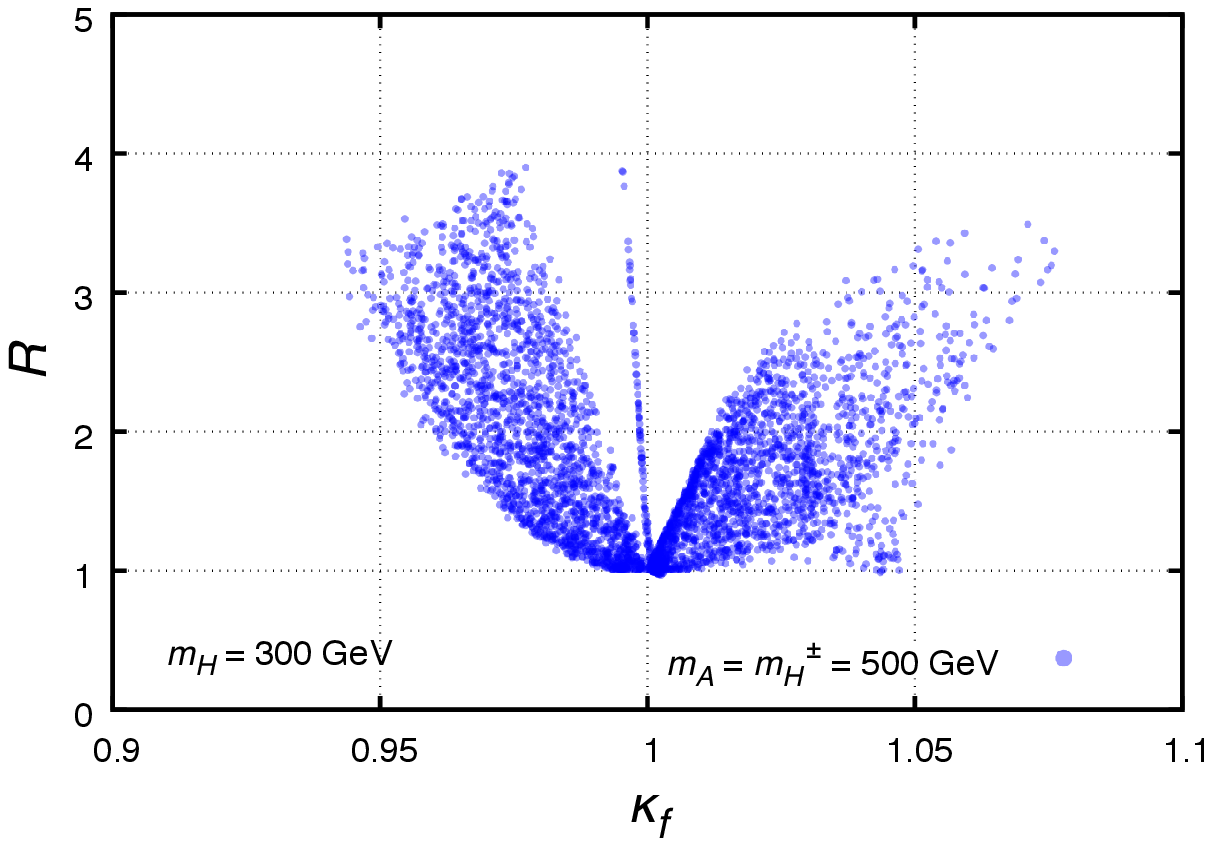}
\caption{Correlation between $\kappa_f^{}$ and $R$ in the case of $m_A = m_{H^\pm} = 300$ GeV (left) and 500 GeV (right). 
In the both figures, we take $m_H = 300\ \text{GeV}$, while $M^2$, $\tan\beta$ and $s_{\beta-\alpha}$ are scanned. 
All the points satisfy the theoretical constraints and experimental bounds explained in this section.}
\label{fig:scan}

\vspace*{\floatsep}

\includegraphics[width=0.45\textwidth]{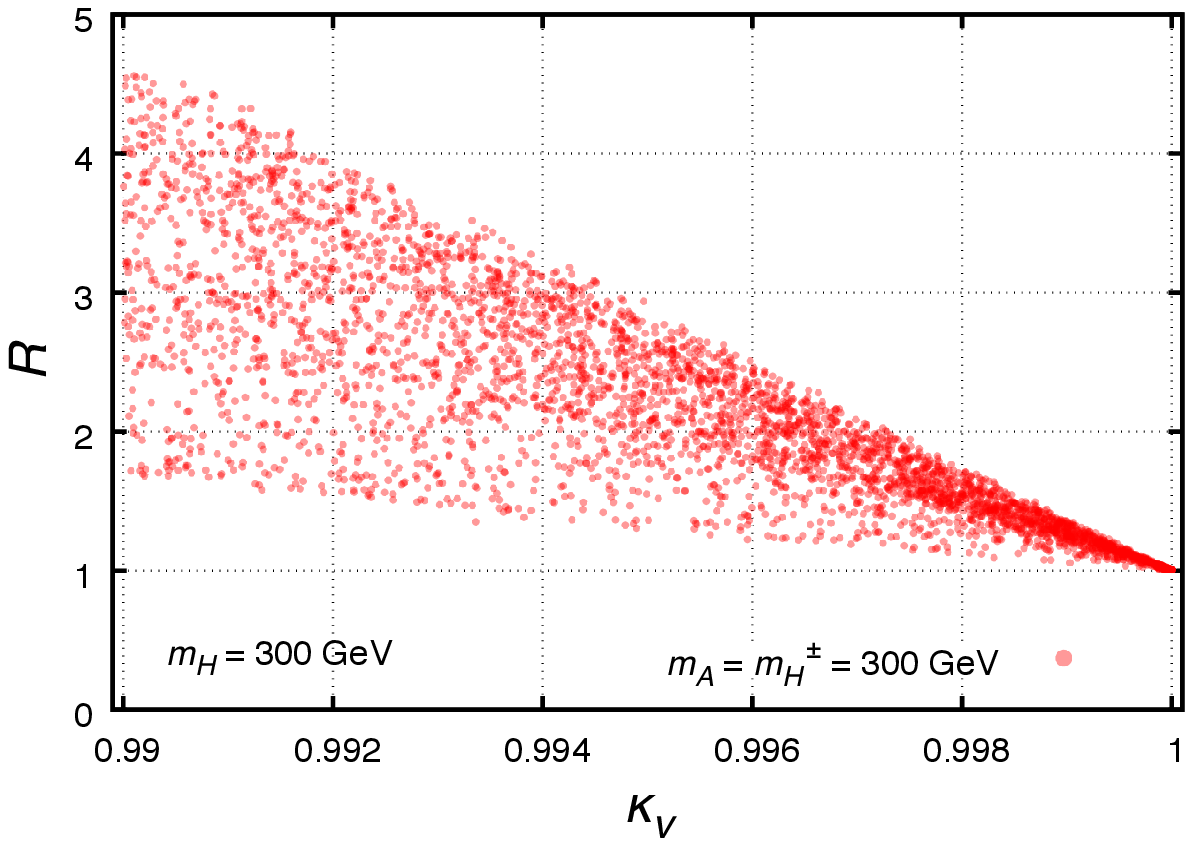}
\includegraphics[width=0.45\textwidth]{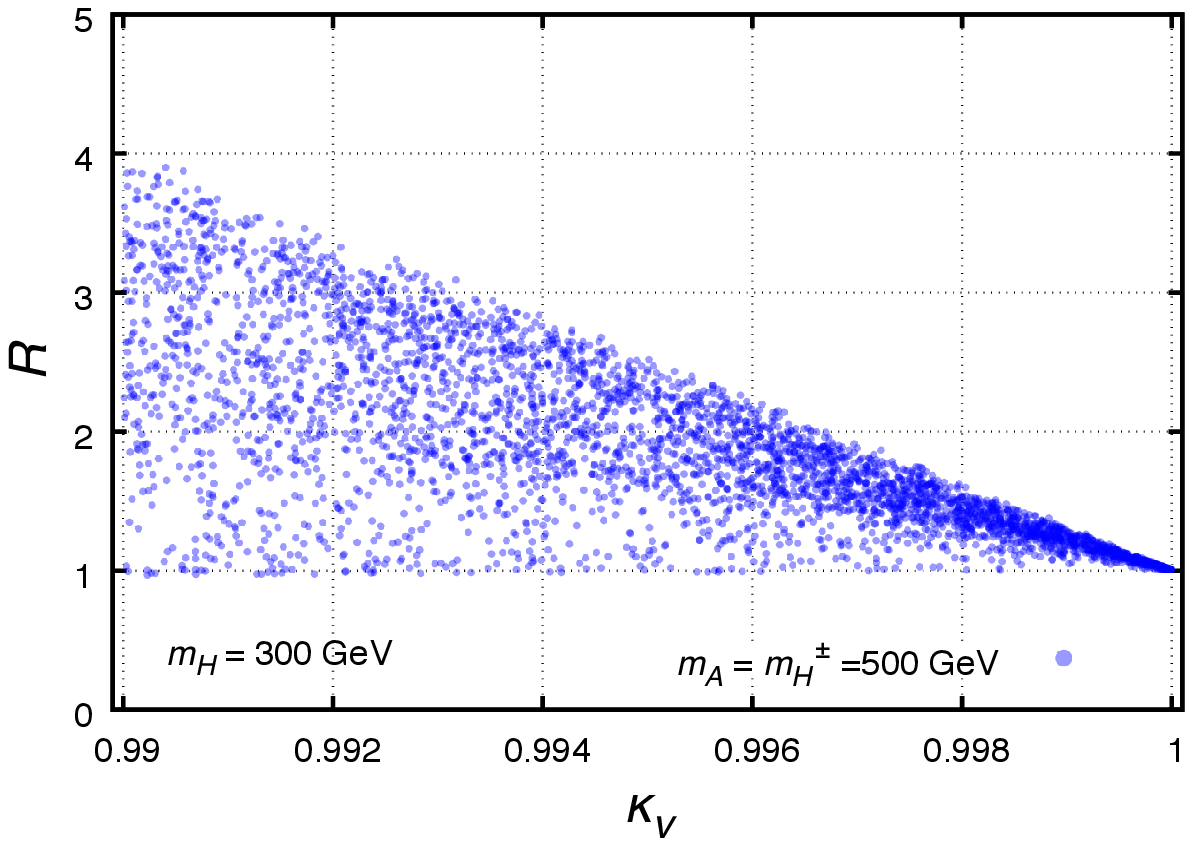}
\caption{Correlation between $\kappa_V^{} (=s_{\beta-\alpha})$ and $R$ in the case of $m_A = m_{H^\pm} = 300$ GeV (left) and 500 GeV (right). 
In the both figures, we take $m_H = 300\ \text{GeV}$, while $M^2$ and $\tan\beta$ are scanned. 
All the points satisfy the theoretical constraints and experimental bounds explained in this section.}
\label{fig:kv}
\end{figure}

In Fig.~\ref{fig:scan}, we scan the value of $s_{\beta-\alpha}$ with $0.99 \leq s_{\beta-\alpha} \leq 1$ under the constraints described in the above. 
The left (right) panel shows the case with $m_A^{} = m_{H^\pm}^{} =300$ (500) GeV, while $m_H^{}$ is fixed to be 300 GeV in the both panels.  
The value of $R$ is distributed in the range between 1 and 4.5 (4) for the case with $m_A^{} = m_{H^\pm}^{} =300$ (500) GeV. 
In Fig.~\ref{fig:kv}, we also show the dependence on the scaling factor of $hVV$ ($V=W,Z$) couplings denoted as $\kappa_V  \equiv g_{hVV}^{\text{THDM}}/g_{hVV}^{\text{SM}} = s_{\beta-\alpha}$. 
It is clarified that the value of $R$ approaches to unity when $\kappa_V^{} \to 1$, because $H$ and $A$ do not appear in the diagram in this limit and all the couplings of $h$ become the 
same value as those in the SM\@. However, once $\kappa_V^{} < 1$ is taken, a significant enhancement is driven by the appearance of the $H$ and $A$ mediations in the diagram depending on the value of $\kappa_f$, 
see also Figs.~\ref{fig:mh300} and~\ref{fig:mA500}.  

\begin{figure}
\centering
\includegraphics[width=0.45\textwidth]{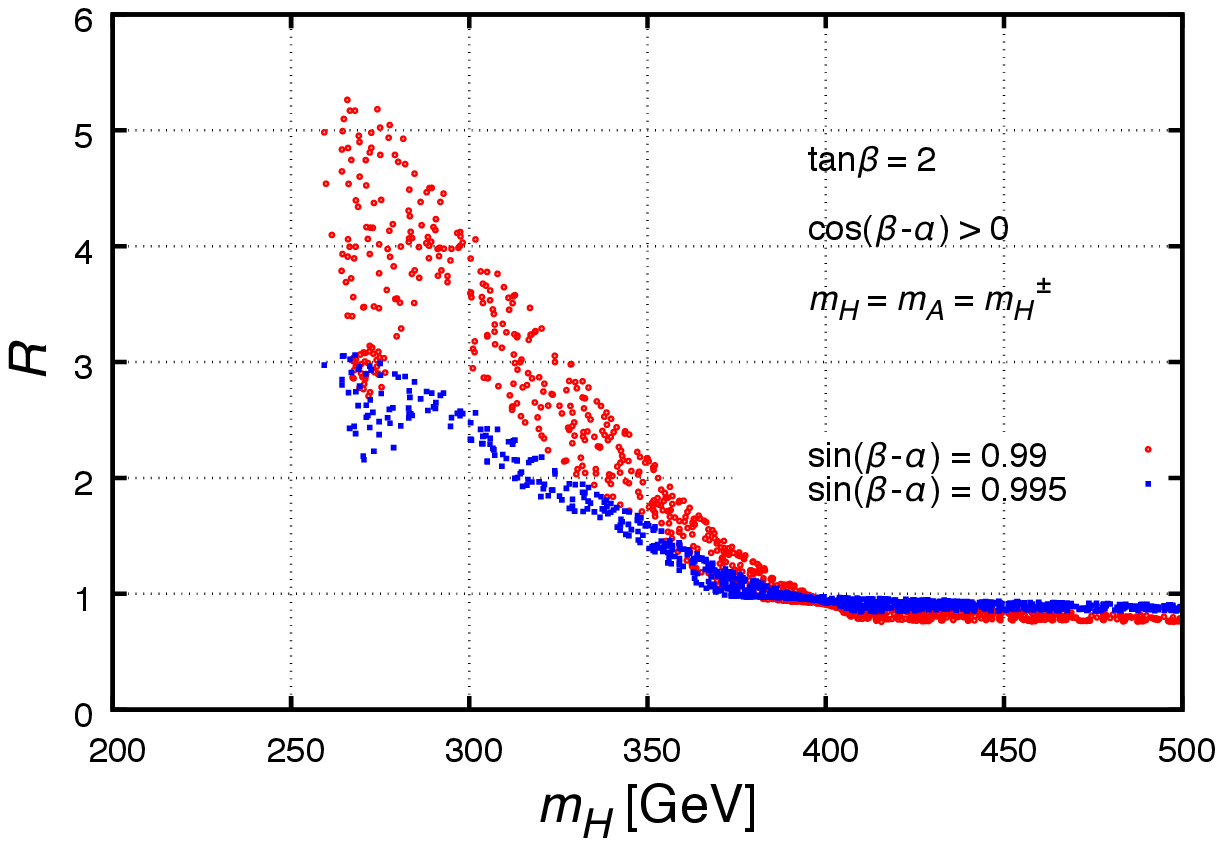}
\includegraphics[width=0.45\textwidth]{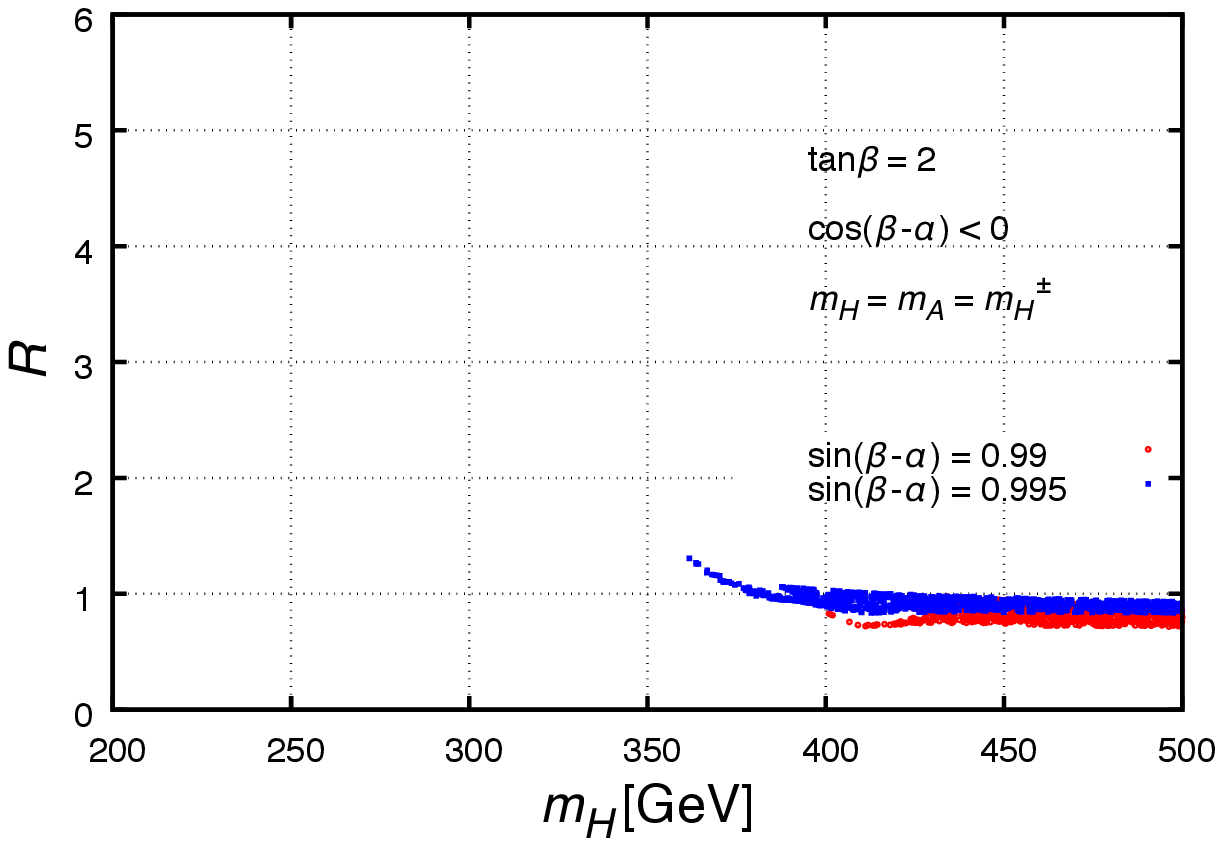} \\
\includegraphics[width=0.45\textwidth]{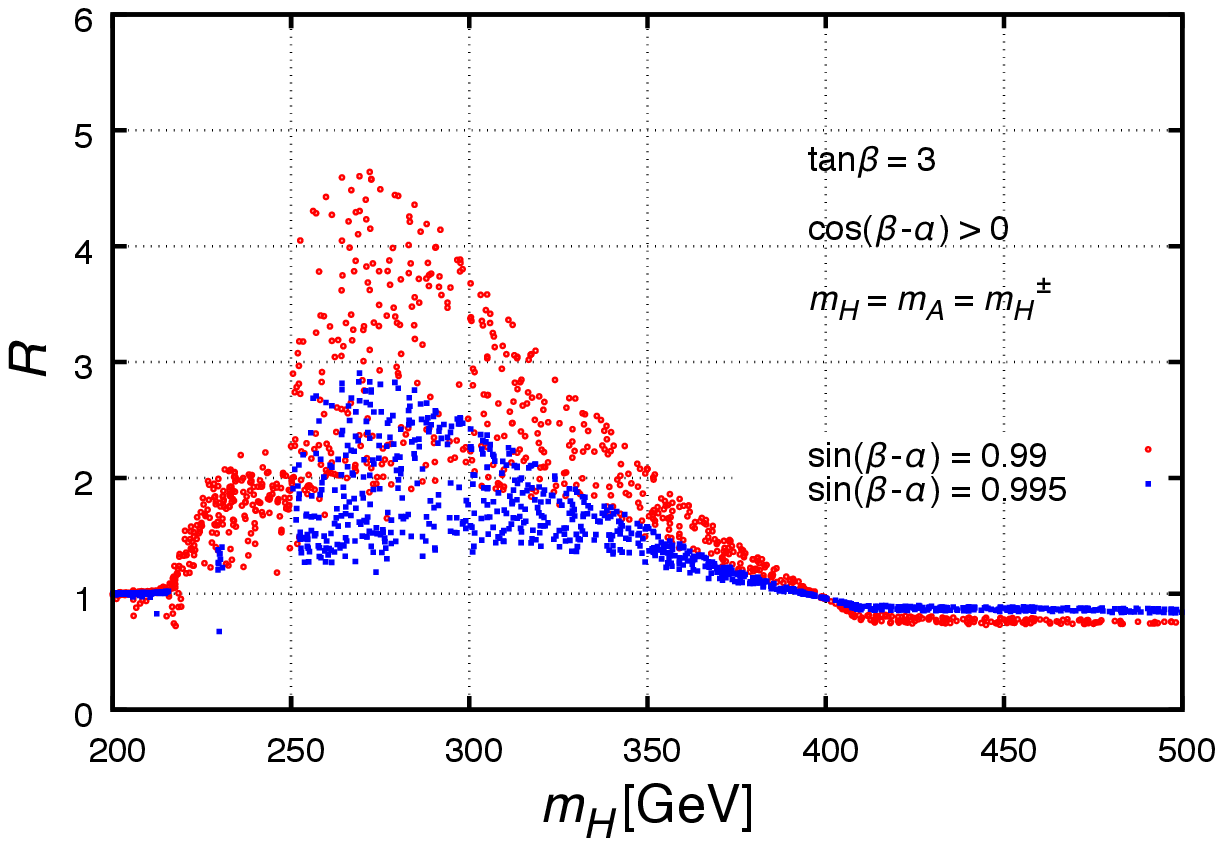}
\includegraphics[width=0.45\textwidth]{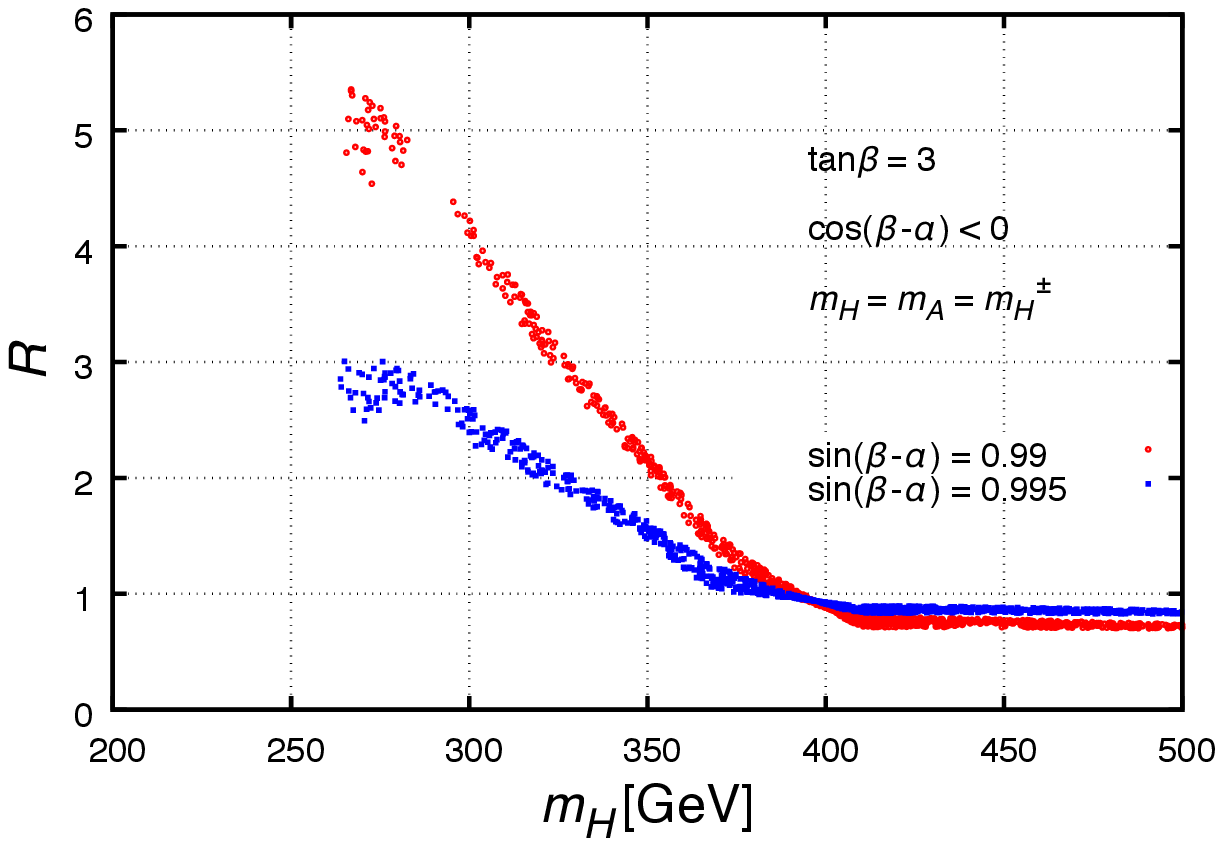} \\
\includegraphics[width=0.45\textwidth]{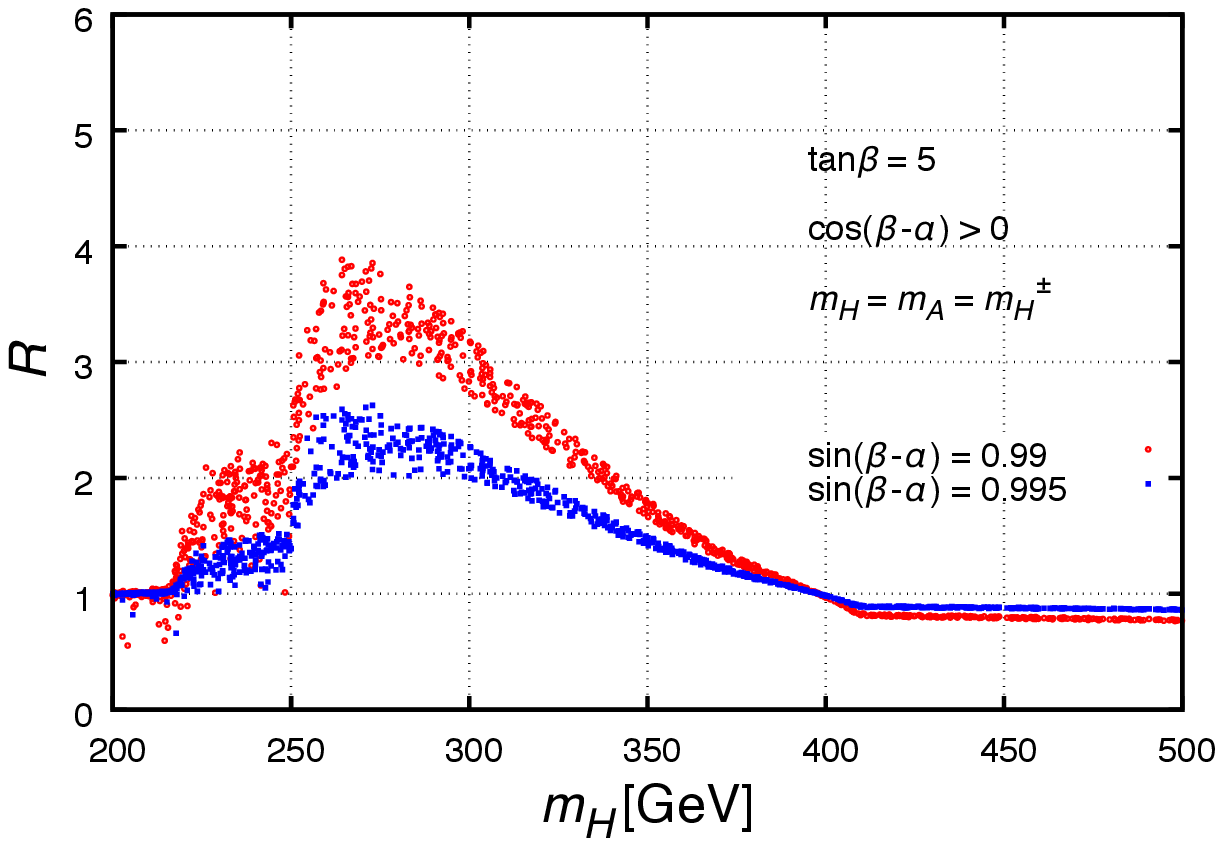}
\includegraphics[width=0.45\textwidth]{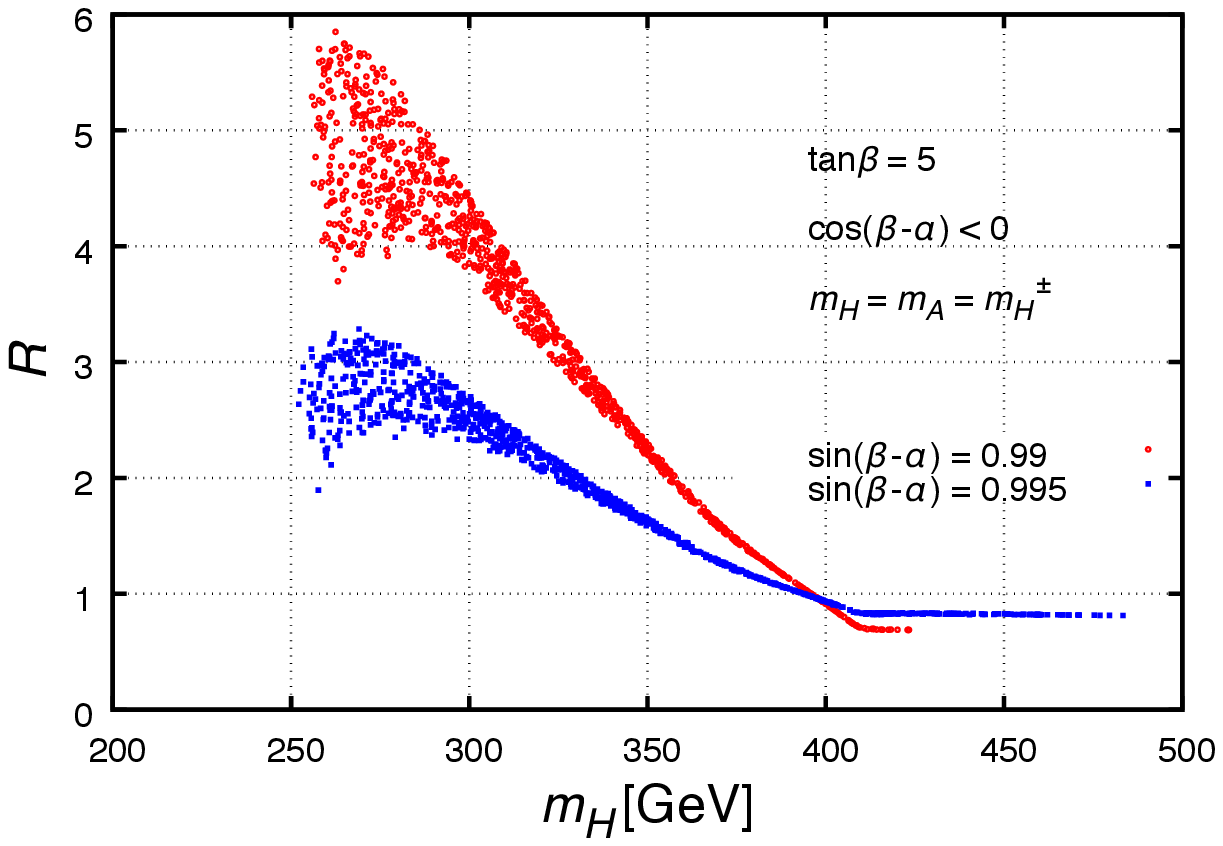}
\caption{Correlation between $m_H^{}\,(=m_A^{} = m_{H^\pm}^{})$ and $R$ for the case of 
$\tan\beta = 2$ (top), $\tan\beta = 3$ (middle) and  $\tan\beta = 5$ (bottom). 
The sign of $c_{\beta-\alpha}$ is taken to positive (left) and negative (right). For all the plots, we take $s_{\beta-\alpha} = 0.99$ (red-circle points) and 0.995 (blue-square points). }
\label{fig:99}
\end{figure}

Finally, we show the mass dependence of the value of $R$ in Fig.~\ref{fig:99} with a fixed value of $\tan\beta$ to be 2 (top), 3 (middle) and 5 (bottom). 
In these plots, we take the degenerate mass of the extra Higgs bosons, i.e., $m_H^{} = m_A^{} = m_{H^\pm}^{}$, and varying the mass range to be from 200 GeV to 500 GeV\@. 
Again, all the points are passed all the constraints explained in the above. 
For the low $\tan\beta$ case such as $\tan\beta = 2$ shown in the top panels, 
the constraint from direct searches excludes the lower mass region, so that the points appear only in the larger mass region.  
For the larger $\tan\beta$ case particularly with $c_{\beta-\alpha} > 0$ (left panels), the lower mass region is allowed, and  
it is clearly seen that the value of $R$ drastically increases at the mass of extra Higgs boson just above 215 and 250 GeV, because of the on-shell $A \to Zh$ and $H \to hh$ decays open. 
We also find that for the larger mass region, e.g., $m_H^{} \gtrsim 400$ GeV the value of $R$ becomes smaller than 1, in which the $H$ and $A$ appearing in the $Zhh$ production are getting off-shell\footnote{In fact, 
for the case with $m_H^{} \sim 400$ GeV, $H$ can still be on-shell as it is produced with the $Z$ boson at the collision energy of 500 GeV.
However, we have checked that the cross section of the $e^+e^- \to ZH \to Zhh$ with $m_H = 400$ GeV is smaller than the other subdominant
contributions such as the $e^+e^- \to  \nu_e \bar{\nu}_e hh$ process, 
because of the phase space suppression. Therefore, we cannot obtain the large enhancement of the cross section at $m_H^{}\sim 400$ GeV or larger. }, so that 
these contributions become unimportant. 
The value $R < 1$ is simply explained by the fact that the contribution containing $h^* \to hh$ as the diagram (a) in Fig.~\ref{diag1} 
becomes smaller than the SM prediction because of the 
smaller $\lambda_{hhh}$ coupling with respect to the SM value, see Eq.~(\ref{lamhhh}).  
We note that in these plots, the value of $\tan\beta$ is fixed, so that the value of $\kappa_f$ is determined to be (1.06,~1.04,~1.02;~0.92,~0.94,~0.96) for $s_{\beta-\alpha} = 0.99$
and (1.04,~1.03,~1.01;~0.95,~0.96,~0.98) for $s_{\beta-\alpha} = 0.995$ where the first (second) three values are the case for $c_{\beta-\alpha}>0~(c_{\beta-\alpha}<0)$ with $\tan\beta =2$, $3$ and 5. 

Before closing this section, we would like to give a brief comment on the difference of the prediction from the other models, 
e.g., models with an isospin singlet scalar field, see for instance~\cite{Lewis:2017dme}. 
In these models, both $\kappa_V\, (V = W,Z)$ and $\kappa_f$ are given by the same factor like $\cos \theta$, where $\theta$ is a mixing angle between 2 CP-even Higgs bosons. 
Therefore, the case ($\kappa_f \sim 1$, but $R > 1$) what we found in the above does not happen, and so our analysis would give an important result to pin down the possible models.  

In addition, we would also like to comment on the possibility of the direct detection of $H$ at future collider experiments such as the HL-LHC\@. 
In the parameter space what we considered in this paper, the branching ratio of $H$ is approximately given by $\text{BR}(H \to hh) + \text{BR}(H \to WW) + \text{BR}(H \to ZZ) \sim 1$, where 
the small portion of the branching ratio is filled by the fermionic decay modes\footnote{For $m_H \gtrsim 350$ GeV, the branching ratio of the $H \to t\bar{t}$ mode can be 10\% level for a smaller value of $\tan\beta$. }. 
Thus, $H$ can be discovered via the $H \to WW/ZZ$ mode. 
In fact at the HL-LHC, there is a study for direct searches of a neutral scalar boson $X$ using the $X \to ZZ$ mode, and the 95\% confidence level upper limit on the cross section ($pp \to X$) times 
the branching ratio $(X \to ZZ)$ is taken in Ref.~\cite{Cepeda:2019klc}. 
This bound can be translated into the constraint on the parameter space in our model, by which smaller values of $\tan\beta$ and/or $m_H^{}$ can be excluded. 
Combining such constraint into our analysis, more restricted results would be obtained.

\section{Conclusions}
\label{sec:conclusions}

We have discussed the correlation between the scaling factor of the Yukawa coupling for the SM like Higgs boson $\kappa_f$ and the 
ratio of the cross section for the $e^+e^- \to hhf\bar{f}$ ($f \neq t$) process normalized to the SM prediction in the Type-I THDM\@. 
We particularly concentrate on the case without the alignment limit, in which resonant effects of the extra neutral Higgs bosons $H$ and $A$ 
provide a sizable enhancement of the cross section and the value of $\kappa_f$ is different from unity at tree level. 
Under the constraints from perturbative unitarity, vacuum stability, electroweak oblique parameters, direct searches for heavy Higgs bosons at collider experiments and 
compatibility of the signal strengths of the discovered Higgs boson, 
we have found that the considerable enhancement of the cross section, typically a few times larger than the SM prediction, can be obtained depending on the value of $\kappa_f$ 
and the masses of the extra Higgs bosons.
The value of $\kappa_f$ is expected to be precisely measured at future collider experiments such as the high-luminosity LHC and the ILC, 
typically with a few percent and one percent level, respectively. 
Therefore, if some deviations in the Higgs boson couplings are found at future colliders, we expect the sizable enhancement of the double Higgs boson production 
and can extract information of the mass of the extra neutral Higgs boson and/or dimensionful model parameter $M^2$ in the Higgs potential. 

Although expected not to give so drastic change, radiative corrections to the
cross-section enhancement studied in this work may be investigated
by using, for example, \texttt{H-COUP}~\cite{Kanemura:2017gbi} to incorporate one-loop
electroweak-corrected vertices or
\texttt{GRACE-loop}~\cite{Belanger:2003sd,Fujimoto:2007bn}
for full one-loop computation.
We leave it for future works.

\begin{acknowledgments}
  All the authors are grateful for fruitful discussions with Masaaki Kuroda, Yoshimasa Kurihara, Kiyoshi Kato,  
  Masato Jimbo, Tadashi Ishikawa, Junpei Fujimoto, Yusaku Kouda and Ryotaro Nara. 

\end{acknowledgments}

\bibliography{references}

\end{document}